# Detection Software Content Failures Using Dynamic Execution Information

Shiyi Kong
School of Reliability and Systems Engineering, Beihang University,
Beijing, China, 100191.
E-mail:buaaksy@buaa.edu.cn

Minyan Lu
School of Reliability and Systems Engineering, Beihang University,
Beijing, China, 100191.
E-mail:lmy@buaa.edu.cn

Bo Sun
System Engineering Research Institute, China State Shipbuilding Corporation Electronics Technology, Co. LTD
Beijing, China, 100036
E-mail: sunbo_bupt@163.com

Jun Ai
School of Reliability and Systems Engineering,
Beihang University,
Beijing, China, 100191.
E-mail:aijun@buaa.edu.cn
Corresponding author

Shuguang Wang
School of Reliability and Systems Engineering,
Beihang University,
Beijing, China, 100191.

*Abstract*—Modern software systems become more and more complex, which makes them difficult to test and validate. Detecting software partial failures in complex systems at runtime assist to handle software unintended behaviors, avoiding catastrophic software failures and improving software runtime availability. These detection techniques aim to find the manifestation of faults before they finally lead to unavoidable failures, thus supporting following runtime fault-tolerant techniques. We review the state-of-the-art articles and find that the content failures account for the majority of all kinds of software failures, but its detection methods are rarely studied. In this work, we propose a novel failure detection indicator based on the software runtime dynamic execution information for software content failures. The runtime information is recorded during software execution, then transformed to a measure named runtime entropy and finally fed into machine-learning models. The machine-learning models are built to classify the intended and unintended behaviors of the objected software systems. A series of controlled experiments on several open-source projects are conducted to prove the feasibility of the method. We also evaluate the accuracy of machine-learning models built in this work.

## I. INTRODUCTION

With the development of computer science and software engineering, software becomes more and more complex. Software reliability has become the core concerns of modern system reliability. Traditional software reliability assurance techniques, like Verification and Validation, are faced with much more challenges when dealing with modern complex software systems. Emergent issues in those systems, including sophisticated interactions between hardware devices and software components, aging problems after long-running, and lacking clear system boundaries, make them more difficult to be tested and supervised[1].

Software Prognostic Health Management (S-PHM) is a kind of runtime software reliability assurance techniques[2]. S-PHM monitors software behaviors and states at runtime, detects and predicts possible failures, and takes proper measures to mitigate failure influence. Detecting errors (partial failures) and then making predictions on the occurrences of the system failures, are the key stages of S-PHM activities[3].

Failures is the deviation of expected functions of the software system[4]. Failure prediction techniques aim to predict the manifestation of failures before they actually occur, as claimed in[5]. Before software systems suffer a failure, it is already in a state in which there is a discrepancy between its actual and the correct internal condition, although the deviation is not perceivable. Such a discrepancy is called an error, which is a partial failure of the whole system[6]. The causes of errors are faults (also called defects). Typical faults are the wrong or missing lines of the code of the applications.

A study conducted by the Software Engineering Institute of Carnegie Mellon University shows that there are still 20% residual defects after software products being released[7].

Resident defects, which are not discovered during the development and testing phases, can cause system failures and unexpected consequences range from a mere minor performance anomaly to a catastrophic accident with not only loss of money and property, but also possible loss of human life[8]. Most resident defects are complex defects that are not easy to be detected during testing phases[6].

These residual defects need to be handled at software runtime to prevent serious damages brought by them. Software online failure detection and prediction methods are proposed to support dealing with the influence brought by residual defects. Existing errors detection and prediction methods can be classified into two categories:

- Log-based methods analyze log files exported from software systems and use text analysis techniques and machine learning algorithms to detect or predict anomalies. The accuracy of these methods depends heavily on the integrity of the log files.

- Symptom-based methods collect runtime dynamic execution information within software applications via instrumentation techniques or system probes. These methods analyze the trace information and detect observable symptoms that can represent anomalies in the traces.

Observable symptoms aforementioned refer to the manifestation of the abnormal software behaviors and status before system failures occur[9]. The symptoms are also called failure mechanisms in some work[10, 11].

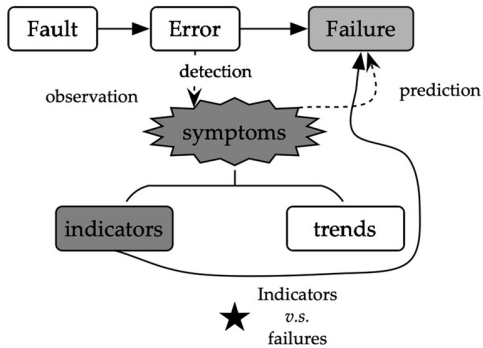

Figure 1 Symptom-based detection methods

Failure mechanisms (the symptoms) can be divided into two parts, indicators, and the trends of them, as shown in Figure 1. For example, memory exhaustion will cause software failures. There is memory usage soaring before software failures. In this failure mechanism, memory usage is the indicator that can be used to predict software failures. The soaring phenomenon is the trends of the indicator.

Avižienis *et.al* in [12] present a taxonomy of software failure types as Figure 2 shows, which has been widely accepted by researchers in this field.

Most software failure indicators in existing research focus

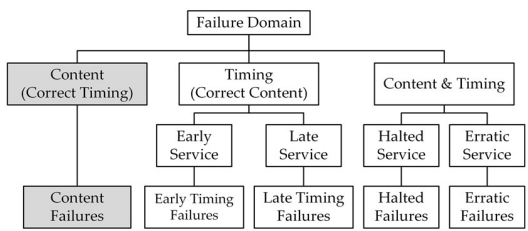

Figure 2 Taxonomy of software failure types.

on non-content failures, especially on aging-related failures (ARFs) [13], hang and crash failures.

However, after investigating a large amount of civil and military software products, researchers in [14] present that software content failures share the largest proportions of all kinds of failures. Content failures refer to *"the content of the information delivered at the service interface (i.e., the service content) deviates from implementing the system function"*, described in [4].

To our best knowledge, there are limited research that focus on content failures.

*Seer* is proposed as a failure prediction approach for software content failures in [5]. Seer uses CPU performance counters as the indicator to predict failures. The accuracy of its classifier model is close to 70%.

Li uses module duration and call times between each two modules as indicators to predict content failures in [11]. Bytecode instrumentation techniques are used to collect runtime data of each function. The accuracy of his classifier model is higher than Seer, above 90%. Instrumentation techniques bring much more accuracy to the prediction as well as runtime overheads.

However, duration and invocation of all modules also bring too many features to the datasets prepared for machine learning and make them sparser, which can cause poor generalization and much more time and resource consumption.

We proposed a novel online detection method for software content failures in this work, using dynamic software execution information. We use Shannon's entropy metrics[15] to measure the dynamic execution information.

Entropy has been proved that satisfy the key properties to make a failure prediction in [13], namely Stability, Monotonically and Integration. A compositive indicator is created to reduce the number of features to improve the generalization while ensuring the accuracy.

The indicator, named software runtime entropy, has been proposed and proven to be of strong relationships with software content failures in this work.

This work has been organized as followings. Section II defines the software runtime entropy measures, Section III introduces the empirical study on the relationships between software runtime entropy and software content failures. Evaluation of the machine learning models is also introduced in Section III. Section IV summaries the major contribution of this work.

## II. DEFINITION OF SOFTWARE RUNTIME ENTROPY

Trace is a series of tuples collected during software execution ordered by time of their occurrence. Execution trace is usually used to study the software execution process [16]. An example of execution trace is shown in Table I. Each line of the data contains 1) a label distinguishing the data collected from the entrance or exit of a function, 2) a function name and 3) a timestamp record current time.

TABEL I SAMPLES OF EXECUTION TRACE

| ID | Label [a] | Function Name | Timestamp [b] |
|---|---|---|---|
| 1 | IN | Main | 10728 |
| 2 | IN | Func A | 10750 |
| 3 | OUT | Func A | 10830 |
| 4 | IN | Func B | 10850 |
| 5 | IN | Func C | 10900 |
| 6 | OUT | Func C | 11000 |
| 7 | OUT | Func B | 11200 |
| 8 | OUT | Main | 11290 |

[a] The label distinguish the data collected at the entrance or exit of a function.
[b] Timestamp presents the data collection time.

### A. Runtime Entropy Definitions

Shannon proposes a definition of entropy to measure an information source. For a given information source piece $X =$

$\{x_1, x_2, \ldots, x_n\}$, the appearance rates of each element $x_i$ is $p_i$. Obviously, there will always be:

$$\sum_{i=1}^{n} p_i = 1. \quad (1)$$

The entropy of this information source can be calculated by Equation 2 as in Shannon's theory:

$$H = -\sum_{i=1}^{n} p_i \log(p_i). \quad (2)$$

Research in [13, 16, 17] have shown that entropy can reflect software execution status. Based on the execution trace collected inner software, we construct the runtime entropy in this section, and use them to depict software execution behaviors and status. There is one parameter in Shannon's formula, the appearance rates $p_i$ of a specific event in the information pieces.

Take a piece of trace data as the information piece. There are two kinds of events defined in this work, function executions and invocations. Event $A_i$ is defined as "the specific function $i$ is on its executing", event $B_{i \to j}$ is defined as "the function $i$ calls the function $j$". The appearance rates of event $A_i$ is denoted as $\alpha_i$, can be calculated as equation 3.

$$\alpha_i = \frac{T_{duration}^i}{\sum_{k=1}^{n} T_{duration}^k}. \quad (3)$$

The software execution time entropy $H_A$ can be calculated as equation 4.

$$H_A = -\sum_{i=1}^{n} \alpha_i \log(\alpha_i). \quad (4)$$

While $T_{duration}^i$ denotes the duration time of a specific function $i$, and can be calculated as equation 5.

$$T_{duration} = T_{out} - T_{in} - T_{subduration}. \quad (5)$$

The following figure 3 shows and example of software duration calculating process.

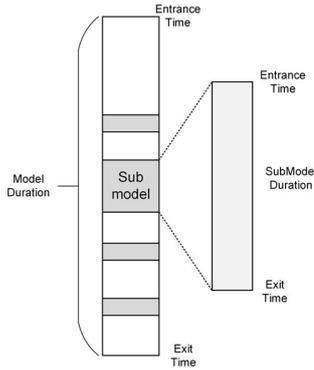

Figure 3 Software duration time.

As for event $B_{i \to j}$, we use function call times and frequency to calculate function call entropy $H_B$.

Figure 4 reveals a sample of function invocations. The nodes represent the functions and the edges represent the invocations. The number on the edge denoted the number of times that the invocation occurs. $i \to j$ denotes the invocation event. $N_{i \to j}$ denotes the number of times that function $i$ calls function j. The $\beta_{i \to j}$ denotes the appearance rates of $B_{i \to j}$ in total call events and can be calculated using equation 6.

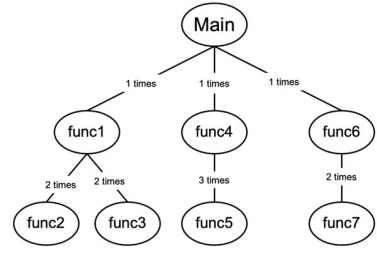

Figure 4 Samples of function invocations.

$$\beta_{i \to j} = \frac{N_{i \to j}}{\sum_{k=1}^{m} \sum_{l=1}^{n} N_{k \to l}}. \quad (6)$$

$m$ means the total number of function callers and $n$ means the total number of functions that have been called.

Then the function call entropy $H_B$ can be calculated using following equation 7.

$$H_B = -\sum_{i=1,j=1}^{m,n} \beta_{i \to j} \log(\beta_{i \to j}). \quad (7)$$

Then the runtime entropy can be calculated using equation 8.

$$H = \frac{1}{2}(H_A + H_B). \quad (8)$$

## III. EMPIRICAL STUDY

In this section, we use a series of empirical study to evaluate the relationships between software runtime entropy and software content failures. Fault injection techniques are used in this process to mimic software failure states. Runtime traces are collected during software execution and then used to calculate runtime entropy. Software runtime entropy accompanied with labels are fed into machine learning model to train a classification model.

*grep*, *flex* and *gzip* are the three open-source software obtained from the *Software-artifact Infrastructure Repository* (*SIR*) used in this paper.

grep is a command-line utility for searching plain-text data sets for lines that match a regular expression. flex (fast lexical analyzer generator) is a free and open-source software alternative to lex. It is a computer program that generates lexical analyzers (also known as "scanners" or "lexers"). gzip is a software application used for file compression and decompression. More details of these three objects can be found in Table 3.

TABLE II BRIEF INTRODUCTIONS OF SOFTWARE OBJECTS

| Name | grep | flex | gzip |
|---|---|---|---|
| **Languages** | C | C | C |
| **Size** | 10068LOC | 10459LOC | 5680LOC |
| **Procedures** | 146 | 163 | 104 |
| **Versions** | 6 | 6 | 6 |
| **Fault seeds** | 18 | 19 | 14 |
| **Test cases** | 470 | 525 | 214 |

## A. Framework

Existing open-source data sets (e.g., MDP from NASA, PROMISE) for software static analysis (e.g., software defect prediction) just consist of static information of software, like software topological structure and static structure measures. So, they cannot be used to do the research on failure detection or failure prediction which need software run-time information.

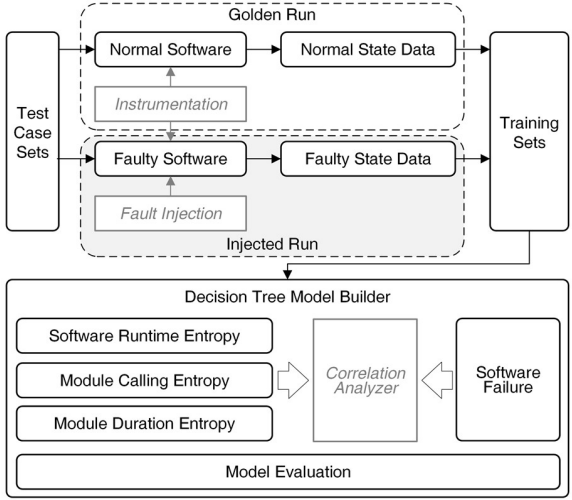

Figure 5 Framework of the overall empirical study.

The lack for run-time data causes those existing symptoms for failure prediction usually come from engineering experience. Like the symptom that more than 90% memory usage will cause failures. The threshold "90%" is set according to engineering experience. Those symptoms are usually used in software performance prediction domains to predict when the resource will be exhausted.

Unlike performance failures usually have performance degradation, software content failures are not so intuitive that we need more information inner software to do the prediction. Content failures are usually not closely related with resource consumption, so we need new indicators for its failure prediction. These indicators should be related with software inner behaviors during execution process.

To obtain software run-time data and then extract the new indicator, we conduct a series of fault injection experiments.

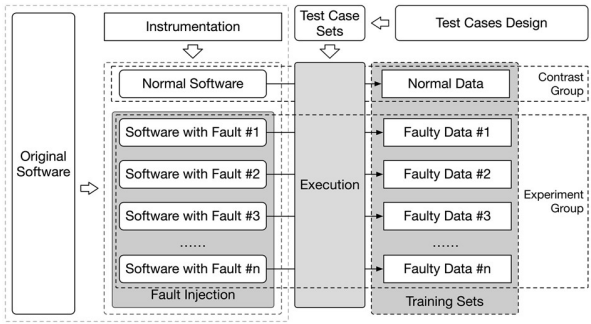

Figure 6 Fault injection process.

The runtime data is fed into the decision tree model to build a classifier for distinguishing failures.

The overall framework of the empirical study is shown in Figure 5. The details of the fault injection process are shown in Figure 6.

Then we process the data and mark them with a label represented if there is a software failure when collecting the data. Using decision tree classification algorithm C4.5, the data obtained from fault injection experiments are analyzed separately.

The data collected from above experiments is imbalanced. The ration between normal execution and failed execution are quite different. This unbalanced data may cause bias when training a classifier. An algorithm named SMOTE (Synthetic Minority Over-Sampling Technique) is introduced to handle this imbalanced. SMOTE will oversample minority class to generate new data for training. Considering reality scenarios, we make the ration between normal execution and failed execution to be 8:2. Due to the data of flex is different with two others, failed execution is far more than normal execution. We use SMOTE twice and make the ration nearly 8:2 and 1:1 separately.

10-fold cross-validation method is chosen to reduce the deviation caused by data selection. The decision tree algorithm owns several parameters. We set confidence factor as 0.25. Confidence factor is used for pruning, smaller confidence factor incurs more pruning. Another important parameter is the minimum numbers of instances per leaf (parameter M). We set it ranges from 2 to 200 and compare the different classification effects of those model in next section.

After obtaining the decision tree models, we need to examine whether the models can truly reflect the relationship between variables involved when building the model (the relationship between software runtime entropy and software failures). Some commonly used measurements in machine learning areas are selected to evaluate the accuracy of these models.

True Positive Rates (TPR), False Positive Rates (FPR), Precisions and F-measure are used to evaluate the classification models.

$$\text{TPR} = \frac{TP}{TP + FN}. \quad (9)$$

While $TP$ denotes the rates that failure samples are classified as failure, $FN$ denotes the rates that normal samples are classified as failure.

$$\text{FPR} = \frac{FP}{FP + TN}. \quad (10)$$

While $FP$ denotes the rates that normal samples are classified as normal, TN denotes the rates that failure samples are classified as normal states.

$$\text{precision} = \frac{TP}{TP + FP}. \quad (11)$$

When calculate F-measure, we use $\beta = 1$ in this work.

$$F - \text{measure} = \frac{(1 + \beta^2)TP \times \text{percision}}{\beta^2 \text{percision} + TP}. \quad (12)$$

The execution times of three software objects are shown in Table 3.

TABLE III EXECUTION TIMES OF THREE SOFTWARE

| Software | Failed | Normal | Total |
|---|---|---|---|
| grep | 1119 | 14227 | 15346 |
| flex | 4141 | 587 | 4728 |
| gzip | 185 | 3025 | 3210 |

The number of instances after SMOTE processing is shown in Table 4.

TABLE IV INSTANCES NUMBER AFTER SMOTE PROCESSING

| Software | Failed | Normal | Total |
|---|---|---|---|
| grep | 4476 | 14227 | 18703 |
| flex-1 | 4141 | 15849 | 19990 |
| flex-2 | 4141 | 5283 | 9424 |
| gzip | 925 | 3025 | 3950 |

The results of the model evaluation are shown in following Table 5, 6 and 7. The runtime entropy indicators of three software objects show a strong correlation with the occurrence of software failures. In grep and gzip, the ratio of normal runs to failure runs is relatively uniform. TPR and FPR of the model are in the normal range and have good classification results. In flex, the failure execution numbers are much larger than normal runs. The unbalanced data make the results have a higher FPR than the other two applications and cause more false alarms.

TABLE V EVALUATION RESULTS OF GREP MODEL

| Parameter M | If SMOTE | Precision | TPR | FPR | F1-measure |
|---|---|---|---|---|---|
| M=2 | No | 0.965 | 0.965 | 0.422 | 0.961 |
| | Yes | 0.933 | 0.932 | 0.106 | 0.932 |
| M=10 | No | 0.963 | 0.963 | 0.458 | 0.957 |
| | Yes | 0.924 | 0.924 | 0.126 | 0.924 |
| M=50 | No | 0.947 | 0.950 | 0.579 | 0.941 |
| | Yes | 0.891 | 0.893 | 0.211 | 0.891 |
| M=100 | No | 0.938 | 0.945 | 0.607 | 0.935 |
| | Yes | 0.876 | 0.888 | 0.231 | 0.877 |
| M=200 | No | 0.933 | 0.942 | 0.663 | 0.929 |
| | Yes | 0.850 | 0.854 | 0.291 | 0.851 |

TABLE VI EVALUATION RESULTS OF FLEX MODEL

| Parameter M | If SMOTE | Precision | TPR | FPR | F1-measure |
|---|---|---|---|---|---|
| M=2 | No | 0.877 | 0.885 | 0.793 | 0.843 |
| | SMOTE-1 | 0.855 | 0.855 | 0.515 | 0.831 |
| | SMOTE-2 | 0.699 | 0.685 | 0.368 | 0.667 |
| M=10 | No | 0.875 | 0.886 | 0.781 | 0.846 |
| | SMOTE-1 | 0.849 | 0.849 | 0.534 | 0.823 |
| | SMOTE-2 | 0.695 | 0.683 | 0.371 | 0.664 |
| M=50 | No | 0.889 | 0.886 | 0.795 | 0.843 |
| | SMOTE-1 | 0.838 | 0.838 | 0.582 | 0.806 |
| | SMOTE-2 | 0.674 | 0.668 | 0.383 | 0.651 |
| M=100 | No | 0.889 | 0.886 | 0.795 | 0.843 |
| | SMOTE-1 | 0.829 | 0.834 | 0.616 | 0.795 |
| | SMOTE-2 | 0.674 | 0.664 | 0.391 | 0.644 |
| M=200 | No | 0.896 | 0.882 | 0.831 | 0.833 |
| | SMOTE-1 | 0.820 | 0.825 | 0.657 | 0.778 |
| | SMOTE-2 | 0.678 | 0.662 | 0.401 | 0.635 |

TABLE VII EVALUATION RESULTS OF GZIP MODEL

| Parameter M | If SMOTE | Precision | TPR | FPR | F1-measure |
|---|---|---|---|---|---|
| M=2 | No | 0.995 | 0.995 | 0.076 | 0.995 |
| | Yes | 0.982 | 0.982 | 0.042 | 0.982 |
| M=10 | No | 0.985 | 0.985 | 0.184 | 0.985 |
| | Yes | 0.970 | 0.971 | 0.065 | 0.970 |
| M=50 | No | 0.969 | 0.971 | 0.362 | 0.969 |
| | Yes | 0.959 | 0.959 | 0.106 | 0.959 |
| M=100 | No | 0.969 | 0.971 | 0.362 | 0.969 |
| | Yes | 0.935 | 0.936 | 0.143 | 0.935 |
| **M=200** | No | - | - | - | - |
| | Yes | 0.916 | 0.912 | 0.268 | 0.906 |

We have plotted the results in following figures (Figure 7, 8 and 9) to reveal some rules of the results and try to explain the influence of parameter M and SMOTE process.

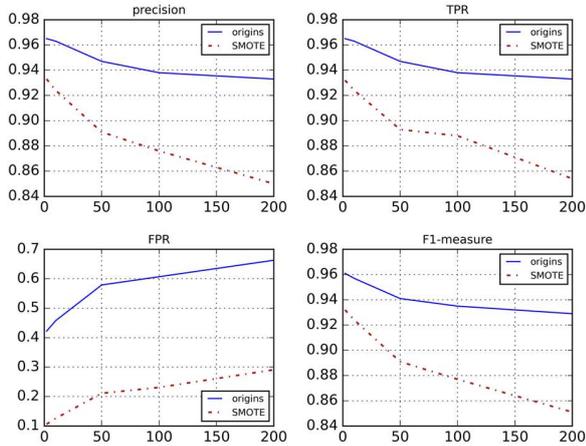

Figure 7 Evaluation results of grep

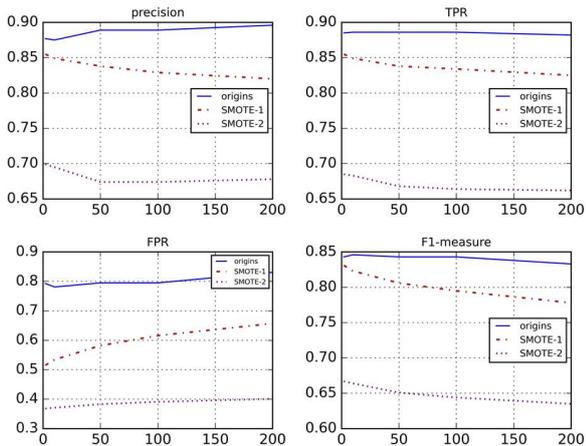

Figure 8 Evaluation results of flex

As with the increase of parameter M, more instances will be classified in the same leaf. there will be less branches and this make the decision clearer. Less rules contains in the model which makes the further prediction work based on the rules simpler. But it also causes the accuracy loss which need a further tradeoff.

When the data is imbalanced, a SMOTE process will produce less false alarming (a lower FPR). Because we generate more failed instance during the SMOTE process. Serious imbalanced data couldn't be corrected well (as shown in the results of flex). On the other hand, SMOTE can also cause a precision loss, more oversampling will cause more accuracy loss.

It shows in the data that more normal instances cause a higher TPR and more failed instances cause a lower FPR. So

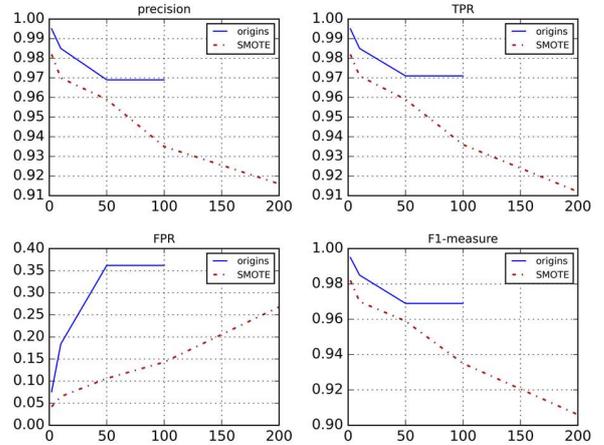

Figure 9 Evaluation results of gzip

under-sampling and over-sampling could be used on the data under different severity.

IV. CONCLUSIONS

A novel indicator for software content failure prediction is proposed in this work which is named software run-time entropy. The indicator has been proved to be with the ability to classify software online health states and can be used to identify software failures with the assist of machine learning models.

A series of fault injection experiments are designed to obtain run-time information to calculate those two indicators. By this way, the issue that lacking for real data has been addressed. Those data are processed and trained with a machine learning algorithm named C4.5. After the training process, we obtain decision tree models and use common measurements (TPR, FPR, precision and F1-measure) to validate the models.

In addition, how to reduce the false alarm rates when there are more failure data than normal data is an interesting issue that needs to be solved in future work.